\documentstyle{article}
\title{
\begin{flushright}
{\bf\normalsize   COLO-HEP-334, LPTHE-Orsay-94/07, SCCS 593}  \\
\end{flushright}
\vskip 10pt
\bf   Quenching 2D Quantum Gravity
}

\author{ {\it C.F. Baillie} \\
         Dept of Computer Science, University of Colorado\\
         Boulder, CO 80309, USA\\ \\
         {\it K.A. Hawick} \\
         Northeast Parallel Architecture Center\\
         Syracuse University\\
         111 College Place\\
         Syracuse, NY 13244, USA\\ \\
         {\it D.A. Johnston}\\
         LPTHE\\
         Universite Paris Sud, Batiment 211\\
         F-91405 Orsay, France$^{1}$\\
}

\begin{document}
\maketitle
\vskip 75pt
                      {\Large
                      \begin{abstract}
%
We simulate the Ising model on a set of fixed random $\phi^3$ graphs, which
corresponds to a {\it quenched} coupling to 2D gravity rather than the
annealed coupling that is usually considered. We investigate
the critical exponents in such a quenched ensemble and compare them
with measurements on dynamical $\phi^3$ graphs, flat lattices and
a single fixed $\phi^3$ graph.
\\ \\
Submitted to Phys Lett B.    \\ \\ \\ \\ \\ \\
$~^{1}$ {\it Address:} Sept. 1993 - 1994, {\it Permanent Address:} Maths Dept,
Heriot-Watt
University, Edinburgh, Scotland \\
%
                        \end{abstract} }
%
  \thispagestyle{empty}
%
%
  \newpage
%
                  \pagenumbering{arabic}
It has recently become possible to calculate the critical exponents for
various spin models on
a particular class of dynamical connectivity lattices using both the methods of
conformal field theory \cite{1} and matrix models \cite{2}. Remarkably,
it is even possible to solve the Ising model exactly in the presence of an
external field on such a lattice, which has proven impossible on fixed 2D
lattices \cite{3}. The key to the calculations, whether in the continuum
conformal field theory formalism or the discrete matrix model formalism,
lies in the observation that putting the model on a dynamical lattice
and allowing it to interact with the lattice is equivalent to coupling the
model
to 2D quantum gravity. This has the effect of ``dressing'' the conformal
weights
$\Delta_0$ of the
critical continuum theory on a flat 2D lattice, to give new conformal
weights $\Delta$
\begin{equation}
\Delta - \Delta_0 = - {\xi^2 \over 2} \Delta ( \Delta - 1),
\label{e00}
\end{equation}
where
\begin{equation}
\xi = - { 1 \over 2 \sqrt{3} } ( \sqrt{ 25 -c } - \sqrt{ 1 -c} )
\label{e01}
\end{equation}
and $c$ is the central charge of the theory in question. This, in turn,
will effect the critical exponents of theory.  These can be calculated from
the conformal weights of the energy and spin operators $\Delta_{\epsilon}$
and $\Delta_{\sigma}$ respectively \cite{3a} which
give $\alpha$ (the specific heat exponent) and $\beta$
(the magnetization exponent)
\begin{eqnarray}
\alpha &=& {1 - 2 \Delta_{\epsilon} \over 1 - \Delta_{\epsilon} } \nonumber \\
\beta &=&  { \Delta_{\sigma} \over 1 - \Delta_{\epsilon} }
\label{e05}
\end{eqnarray}
in the usual fashion.
Given $\alpha$ and $\beta$ from the conformal field
theory we can then use the various standard scaling relations
to calculate the full array of exponents. For the exactly soluble case of the
Ising
model on a dynamical lattice the standard scaling relations
can be shown to hold, so it is not too great a leap of faith to assume their
validity for other models.

There is a growing body of numerical evidence
confirming the new dynamical exponents in simulations \cite{4a,4,4b}, both on
dynamical
triangulations and their dual dynamical $\phi^3$ graphs with both spherical
and toroidal topology lattices \footnote{The topology of the underlying lattice
is not
expected to affect the spin model critical exponents, though it does have an
effect
on the exponent $\gamma_{string}$.}. The matrix model approach suggests that
any
dynamical polygonalization of a surface should give the same exponents
for a given model, so there should be universality in this sense.
If we
think of the numerical simulations purely as an exercise in statistical
mechanics
it is clear they represent taking an annealed average over the various random
graphs
on which the spin models live. The partition function
of the Ising model on dynamical lattices, for instance, is
\begin{equation}
Z = \sum_{G} \sum_{\{ \sigma \}} \exp \beta \left( \sum_{<ij>} G_{ij} \sigma_i
\sigma_j
+ H \sum_i \sigma_i \right)
\label{e07}
\end{equation}
where the connectivity matrix of the lattice is $G_{ij}$.
The dynamics of the lattice, represented by the sum over random $\phi^3$ graphs
or their dual triangulations
$\sum_{G}$, takes place on the same timescale as the dynamics of the spins,
which is manifest in the numerical simulations where the local ``flip'' moves
that change
the lattice connectivity to perform the sum $\sum_{G}$
are mixed in with spin cluster updates. An
interesting related question, to which the answer is not immediately
clear, is what exponents would be measured in a simulation on
a {\it fixed} 2D quantum gravity graph.

A first reflex is to guess that with a fixed lattice we would find
the Onsager exponents. This appears to be correct when we consider
random lattices with ``weak'' disorder - ie sufficiently close to regular
2D lattices, where the Harris criterion (roughly paraphrased as
``weak disorder has no effect if $\alpha<0$'') gives the Ising model
as a marginal case, and simulations come down on the side of Onsager
exponents \cite{6} \footnote{The
analytical work of the Dotsenkos \cite{Dot}, however,
suggests that randomness in the bond {\it values}
can weaken
the specific heat singularity.}. Other fixed random lattices
which are close in some sense to flat lattices \footnote{Having a fractal
dimension of 2, for instance.}, such as Voronoi lattices
also give Onsager exponents \cite{7}. There has been one simulation to date
on a single random $\phi^3$ graph \cite{8} generated by the Tutte algorithm
which proved an efficient means of producing large pure
(ie no matter) 2D gravity lattices \cite{9}, that found results that appeared
to be compatible with the Onsager exponents.  A moment's reflection suggests
that this is a rather surprising result - the fractal dimension of
2D gravity coupled to conformal
matter with central charge $c$ has been calculated \cite{10}, and modulo some
caveats
about lattice dependent effects on the lower moments of the density
of nodes \cite{11}, is given by
\begin{equation}
D = \left( 1 + { \sqrt{ 13 -c } \over \sqrt{ 25 -c }} \right) {\gamma Q \over
2}
\label{e08}
\end{equation}
where $Q= \sqrt{(25 - c) / 3}$ and $\gamma = (Q - \sqrt{ Q^2 - 8}) /
2$. We find that $D$ for $c=1 / 2$, the Ising model interacting
with a dynamical lattice, is equal to 3 and $D$ for $c=0$, ie pure
2D gravity, is $(5 + \sqrt{13}) / 3 \simeq 2.8685..$. We would thus expect
to find $d \simeq 2.8685$ for the graph used in \cite{8}, which is
very different from the 2 of a flat lattice. Unfortunately,
we can't deal analytically
with a single fixed $\phi^3$ graph to see what exponents we might
expect, but we can aspire to taking a {\it quenched} average over
a set of different fixed $\phi^3$ graphs.

It is often the case in statistical mechanics that quenched averages, in which
the disorder is frozen in on the timescale of the spin dynamics, are more
interesting
than annealed averages, particularly in systems where frustration is present
such as
spin-glasses \cite{5a}.
All of the couplings are positive in the model of equ.(\ref{e07}), so we have
no
frustration and hence would not expect to see any of the rich behaviour of
\cite{5a},
but it is still interesting to ask what happens to the critical exponents of
the
model when we consider a quenched sum over the random graphs rather than an
annealed
sum. In a simulation we would implement such a quenched summation by simulating
our Ising model on an appropriate set of fixed random graphs and then taking
a cross lattice average to obtain our observables, as is done in the
simulation of spin glasses to average over different sets of spin couplings.
In \cite{12} it was suggested that applying the replica trick \cite{5a}
in conjunction with the continuum methods of \cite{1} allowed the
calculation of the exponents in a quenched average over 2D gravity
graphs. For a quenched summation the free energy $F$ is calculated
on each graph before
summing over the random graphs
\begin{equation}
F = \sum_{G} \log Z (G) ,\nonumber \\
\label{e14}
\end{equation}
where $Z(G)$ is  the partition function for the Ising
model on a particular random graph
\begin{equation}
Z(G) =  \sum_{\{ \sigma \}} \exp \beta \left( \sum_{<ij>} G_{ij} \sigma_i
\sigma_j
+ H \sum_i \sigma_i \right).
\end{equation}
The awkward logarithm inside the sum can be dealt with by
replacing $\log Z \simeq (Z^n - 1)/n$, $n \rightarrow 0$ to
give a more familiar partition function, now with $n$
replicas of the original matter. Observables are
calculated with this replica partition function and the limit $n
\rightarrow 0$ is taken in order to obtain the desired quenched
averages. Translating this approach directly to the continuum language
of \cite{1}, we can obtain the quenched value of
the one point function $F_{\Phi}(A)$ of an operator $\Phi$, the
magnetization or the energy for instance,
by inserting one copy $\Phi^a$ of $\Phi$ into the
fixed intrinsic area partition function $Z_n(A)$ for $n$ copies
of the matter theory
\begin{eqnarray}
F_{\Phi}(A) &=& { \int D \phi D X \exp ( -S^n ) \delta \left( \int e^{\xi \phi}
\sqrt{ \hat g} d^2 z - A \right)
\int \Phi^a e^{\rho \phi} \sqrt{\hat g} d^2 z \over Z_n(A) } \nonumber \\
&n \rightarrow 0&
\label{e18}
\end{eqnarray}
where the action $S^n$ is composed of a Liouville part and
a Feigen-Fuchs action \cite{13} for $n$ matter copies
\begin{equation}
S^n_M = { 1 \over 2 \pi} \sum_{a=1}^n \int d^2 z \left( \partial X^a \bar
\partial X^a + { i \over 2} \alpha_0 \sqrt{\hat g}
\hat R X^a \right)
\label{e19}
\end{equation}
and $\alpha_0$ is determined by $c = 1 - 12 \alpha_0^2$. Applying the standard
scaling arguments of DDK \cite{1} to such a theory and taking the limit $n
\rightarrow
0$ to get the quenched averages, we find the following values for
the quenched conformal weights
\begin{equation}
\Delta_{quenched} = { \pm \sqrt{1  + 24 \Delta_0} - 1  \over 4}.
\label{e21}
\end{equation}

If we plug in the values for the energy and spin operators in the Ising model
we find the following quenched weights
\begin{eqnarray}
\Delta_{\epsilon} &=& {\sqrt{13} -1 \over 4} \simeq 0.6513878 \nonumber \\
\Delta_{\sigma}  &=& { \sqrt{5/2} - 1 \over 4 } \simeq 0.1452847,
\label{e23}
\end{eqnarray}
which, using equ.(\ref{e05}), gives the following values for the critical
exponents
$\alpha$ and $\beta$
\begin{eqnarray}
\alpha &=&  {6 - 2 \sqrt{13} \over 5 - \sqrt{13} } \simeq -0.8685170 \nonumber
\\
\beta &=&  { \sqrt{5/2} - 1 \over 5 - \sqrt{13}} \simeq 0.4167516.
\label{e24}
\end{eqnarray}
Applying the scaling relations then gives the full set of quenched critical
exponents,
which we repeat from \cite{12} in Table.1 below for completeness, along with
the fixed lattice (Onsager) and annealed (dynamical lattice KPZ/DDK) exponents.

\begin{center}
\begin{tabular}{|c|c|c|c|c|c|c|c|c|c|c|c|c|c|c|c|c|c|c|} \hline
$Type$& $\alpha$  & $\beta$  & $\gamma$ & $\delta$ & $\nu $ & $\eta $\\[.05in]
\hline
$Fixed$& $0$  & $\frac{1}{8}$ & $\frac{7}{4}$ & $15$ & $1$   &
$\frac{1}{4}$\\[.05in]
\hline
$Annealed$& $-1$  & $\frac{1}{2}$ & $2$ & $5$ & $\frac{3}{d}$   & $2 -
\frac{2d}{3}$\\[.05in]
\hline
$Quenched$& $-0.8685169$  & $0.4167516$ & $2.0350137$ &  $5.8830375$ &
$\frac{2.8685169}{d}$     & $2 - 0.7094306d$\\[.05in]
\hline
\end{tabular}
\end{center}
\vspace{.1in}

\centerline{Table 1:  Critical exponents for Ising models}

\bigskip
We thus predict a set of quenched exponents that are close, but not equal to,
the dynamical lattice exponents. The non-rational exponents might be a little
disquieting at first site, but they seem to be a consequence of forcing
the model to live in a gravitational background with the ``wrong'' value
of $c$ (ie $c = 0$ effectively as we have a total
central charge of $nc, \ n \rightarrow 0$ in the quenched theory rather
than the natural $c = 1 / 2$). Similar non-rational exponents were calculated
and observed for percolation, where the natural value of $c$
is eaual to zero,
on $c \ne 0$ string worldsheets in \cite{14}. In fact if we insert the
appropriate values
for the fractal dimension calculated from equ.(\ref{e08}) into the above table
we see that we obtain the same values (1, 0 respectively) for the
exponents $\nu$ and $\eta$ in both the annealed and quenched cases.
If we take these two exponents as our starting point
and calculate the rest from the scaling relations,
we could argue that the quenched model differs
only from the annealed model in living in a slightly different dimension.
We have not found the fixed lattice values for the exponents in the above
calculation
because switching off 2D gravity completely corresponds to the limit $c
\rightarrow - \infty$
rather than $c \rightarrow 0$. The effects of 2D gravity are thus still
visible,
even in an ensemble of fixed $\phi^3$ graphs.

So much for the theory. The results of \cite{8} suggest that
the Onsager exponents appear on a fixed $\phi^3$ graph, so an average over
a set of such exponents would still give Onsager exponents,
in spite of the above arguments. In this paper we report
on a simulation where we implement a quenched average explicitly
by simulating $O(100)$ Ising models on different fixed $\phi^3$ graphs
simultaneously
using a parallel computer. We have resurrected
and modified the program used in earlier simulations
on dynamical lattices \cite{4a} for this purpose, so we
perform a microcanonical (fixed number of nodes) simulation on $\phi^3$ graphs
of spherical topology without tadpoles or self-energy bubbles
with $N=124,250,500,1000$ and $2000$ nodes. We used the Tutte
algorithm to generate the one hundred or so $\phi^3$ graphs
of each size that were required. We
simulated at a range of $\beta$ values from $0.1$ to $1.1$, using hot and cold
starts
where appropriate, with 10K metropolis thermalization sweeps and 50K Wolff
cluster algorithm update sweeps at each $\beta$ value and graph size. Note that
the absence of $NFLIP = N$ dynamical flip moves between each spin update
in the quenched simulation means that our statistics are rather more modest
than the dynamical lattice statistics in \cite{4a}. The various observables
were measured
on each graph and a cross graph average taken to implement the quenched sum.
The errors quoted are those coming from the average across the different
graphs, rather than individual graph statistical errors.

The standard observables were
measured: the energy $E$, specific heat $C$, magnetization $M$, susceptibility
$\chi$,
correlation length $\xi$ and Binder's cumulant for the magnetization
\begin{equation}
U = 1 - { <M^4> \over 3 <M^2>},
\end{equation}
along with the acceptance for the Wolff algorithm, to verify that an
appropriate
number of sweeps was being carried out. The number of Wolff sweeps was
adjusted so that (1 / average size of Wolff cluster) sweeps were performed per
measurement at each $\beta$. In our original dynamical lattice
simulations in \cite{4a} we had some difficulty in finding a reliable
estimate for the crossing point of Binder's cumulant
for different lattice sizes, which is the
usual means of locating the critical point with these methods, so we
adopted an alternative approach that gave good results for the
exactly soluble case of the Ising model, where the critical point is
known analytically. The expected scaling behaviour for Binder's
cumulant is $U \simeq U( t N^{1 / \nu D} )$, so its derivative with
respect to $t$, or equivalently $\beta$, will scale as $1 / \nu D$ at
$\beta_c$. We would also expect the {\it maximum} slope to scale in the same
way
\begin{equation}
\max \left( { d U \over d \beta} \right) \ \simeq \ N^{1 / \nu D}
\end{equation}
which can be used to obtain $\nu D$. Knowing this the finite size scaling
relation
\begin{equation}
| \beta_c^N - \beta_c^{\infty} | \ \simeq  \ N^{ - 1 / \nu D}
\label{fss}
\end{equation}
with $\beta_c^N$ obtained from either the peak in the specific heat or the
maximum
slope in the cumulant, can be used to extract $\beta_c^{\infty}$.
In Fig. 1 we show the maximum slope of the cumulant plotted against
$N$ on a logarithmic scale along with the fitted line
with slope $1 / \nu D = 0.32(1)$. We thus have $\nu D = 3.1(1)$, which can
then be fed into the finite size scaling relation in equ.(\ref{fss})
to get $\beta_c^{\infty}$. In Fig. 2 we show the results for both the maximum
slope in the cumulant and the peak in the specific heat, giving
$\beta_c^{\infty} = 0.778(2)$, where we have dropped
the smallest lattice size from the fit to the cumulant slope.
We also found that it {\it was} possible to extract $\beta_c$ from
the crossing of cumulants for different lattice sizes with the
current data, which gives an additional estimate of $\beta_c=0.78(1)$
compatible with the value above, as is the corresponding estimate of $\nu D$
with this method.

The scaling relation $\alpha = 2 - \nu D$ thus gives $\alpha = - 1.1(1)$ for
the specific heat critical exponent, which is a long way from the
logarithmic scaling ($\alpha=0$) given by the Onsager exponents. It is also
possible to attempt to extract $\alpha$ directly from the singular
behaviour of the specific heat
\begin{equation}
C \ \simeq \ B + C_0 \ t^{- \alpha}
\end{equation}
or from finite size scaling
\begin{equation}
C \ \simeq \ B' + C'_0  \ N^{\alpha / \nu D}.
\end{equation}
Using the
singular behaviour proves to be hopeless with our data as essentially any value
for $\alpha$ can be obtained with comparable $\chi^2$s, but the finite
size scaling fit gives a minimum $\chi^2$-fit for $B' = 1.8$
and $\alpha / \nu D = - 0.37(1)$, which gives $\alpha = - 1.1(1)$,
in good agreement with the value obtained indirectly from the scaling
relation.

A measurement of the exponent $\gamma$ governing the divergence
of the susceptibility is less useful in distinguishing
between the different sets of exponents as the quenched and
annealed values are essentially identical ($\simeq 2$)
and the Onsager value is close to these (1.75).
A fit to
\begin{equation}
\chi = \chi_0 \ t^{-\gamma}
\end{equation}
gives $\gamma = 2.1(2)$ on the $N=2000$ lattices
and $\gamma = 2.0(2)$ for $N=1000$, which
although it favours the dynamical/quenched values,
is not really sufficient to  exclude the Onsager values.
The finite size scaling fit
gives a much poorer fit with an even larger value of $\gamma$. In Fig. 3 we
plot
the divergence in $\chi$ for the two lattice sizes along
with lines of slope 1.75 and 2. It is clear from the diagram
that it is indeed not really possible to decide conclusively
between the candidate values from the data.
Fitting $\beta$ is also rather unilluminating, whether by using
$M \simeq M_0 \ t^{\beta}$ or $M \simeq M_0' \ N^{- 1 / \nu D}$ as we find
values of 0.2-0.3 stranded in no man's land between Onsager
and dynamical/quenched exponents. We found in \cite{4a} that the fits
to $\beta$ on dynamical lattices were the poorest, so a similar
result here is not surprising. It is not clear why the fit to the magnetic
exponent should be so poor, but it was pointed out in \cite{4b}
that the dynamical critical exponent $z$ for the magnetization
was much larger than that for the energy on dynamical lattices
(with the Wolff algorithm) possibly because of clusters becoming
trapped in ``bottlenecks'' in the graph geometry. It is therefore
conceivable that one has to work harder to overcome autocorrelation
and finite size effects in the measurements of $\beta$ than for
other exponents, whether on dynamical or quenched 2D gravity graphs.

We can also fit the exponent $\nu$ by looking at the divergence
of the correlation length $\xi$ at $\beta_c$. As this involves
a double fit, first for the correlation length
\begin{equation}
C_{ij} \ =  \ < { 1 \over n (r) } \sum_{ij} \sigma_i \sigma_j  \delta ( d_{ij}
-
r)> \  \
\simeq \    \exp ( - { d_{ij} \over \xi} )
\end{equation}
(where $d_{ij}$ is the geodesic distance on the lattice
between points $i$ and $j$ and
$n(r) = \sum_{ij} \delta ( d_{ij} - r)$) and then
to the divergence in $\xi$, the fit is not likely to be very accurate.
Doing this however, we find $\nu = 0.87(2)$
for the data on the $N=2000$ graphs,
which is identical to value of $\nu$ we measured on dynamical graphs in
\cite{4a}.
If we take the theoretical values for $D$ for both the quenched and annealed
graphs
we would expect $\nu =1$, just as for the Onsager exponents.

In summary, the fits described above are thus compatible with both those in
\cite{4a}
on dynamical $\phi^3$ graphs
and with the similar quenched exponents. The best results are
for $\nu D$ and $\alpha /  \nu D$, both of which give a cusp ($\alpha = -
1.1(1)$)
in the specific heat, rather than a logarithmic divergence, as expected for
a flat 2D lattice. The measured value of $\gamma$
on the largest lattices used, 2.1, is close to the $ \simeq 2$
expected on dynamical/quenched graphs, but the Onsager value of 1.75
lies not far outside the
error bars. The measured value of $\nu$ is identical to that found
for dynamical graphs, but one cannot use $\nu$ to distinguish between Onsager
and dynamical exponents. The values of $\beta$ measured are as disastrous
as those found in \cite{4a} on dynamical lattices, and hence not very useful.
Finally, it is worth noting that the measured value of $\beta_c^{\infty}$ in
the
quenched simulation is close to the calculable $\beta_c$ on dynamical
graphs (0.7733185).

Although it would be dangerous to claim that there was conclusive evidence for
non-Onsager
exponents in the quenched ensemble on the basis of fits to $\nu D$ and $\alpha$
alone,
the data for these is at least consistent with this, and none of the other fits
contradict it.
Unfortunately, the results are nowhere near accurate enough to
distinguish between the postulated
quenched exponents and dynamical graph exponents, even with the $O(100)$
different
graphs used. If we accept that the quenched ensemble {\it does} give
non-Onsager
exponents we are left with the task of explaining the results of \cite{8}
on a single 2D gravity graph, where acceptable fits were found to the Onsager
values. It is possible that taking the continuum limit in two different
ways, on a fixed graph and in an ensemble of fixed graphs, simply gives
different exponents,
though as one expects self-averaging
for quantities such as the energy
and magnetization it is
difficult to see how this might happen. Another possibility is that the
particular random
graph chosen in \cite{8}
produced a set of pseudo-Onsager exponents by chance,
as there appears to be a wide variation in the
apparent critical points and exponents if these are fitted
in the individual graphs in our ensemble.
Earlier simulations on fixed random lattices, of the
XY model in particular \cite{Wh}, have shown that
finite size effects can be particularly important too.
In closing it is amusing to note that, formally at any rate, we would expect to
see
the same quenched exponents appear in a simulation of an Ising model
on a {\it single} dynamical graph in which the back reaction of the spins
on the lattice was switched off (by
always carrying out flips
rather than performing a Metropolis test
on the resulting energy change) as this also has $c=0$.

This work was supported in part by NATO
collaborative research grant CRG910091.
CFB is supported by DOE under
contract DE-FG02-91ER40672, by NSF Grand Challenge Applications
Group Grant ASC-9217394 and by NASA HPCC Group Grant NAG5-2218.
DAJ is supported at LPTHE
by an EEC HCM fellowship, EEC HCM network grant and an Alliance grant.
Some of the analysis of Binder's cumulant
was carried out using programs written by Andre
Krzywicki, whom DAJ would also like to thank for useful discussions.
The simulations reported in this work were carried out on the
Edinburgh Concurrent Supercomputer (a Meiko Computing Surface) while
one of us (KAH) was employed by the Edinburgh Parallel Computing
Centre.  These simulations used in excess of 32,000 T800
transputer-hours of computation time.

\vfill
\eject

\vfill
\eject
\centerline{\bf Figure Captions}
\begin{description}
\item[Fig. 1.]  The maximum in the slope of Binder's
cumulant is plotted against the lattice size to extract
$\nu D$. The best fit is shown as a line.
\item[Fig. 2.] The $\beta$ value of the maximum in the slope of Binder's
cumulant and the peak in the specific heat are plotted against $N^{- 1 /
\nu D}$ to extract the critical coupling $\beta_c^{\infty}$, given by
the intersection of the best fit lines with the $y$ axis.
\item[Fig. 3.] A finite size scaling plot of $\chi$, with lines
of slope 1.75 (the steeper) and 2 added for comparison.
\end{description}

\begin{thebibliography}{99}
\bibitem{1} V.G. Knizhnik, A.M. Polyakov and A.B. Zamolodchikov, Mod. Phys.
Lett. {\bf A3} (1988) 819;\\
            F. David, Mod. Phys. Lett. {\bf A3} (1988) 1651;\\
            J. Distler and H. Kawai, Nucl. Phys. {\bf B321} (1989) 509.
\bibitem{2} E. Brezin, C. Itzykson, G. Parisi and J.B. Zuber, Commun. Math.
Phys. {\bf 59} (1978) 35;\\
            M.L. Mehta, Commun. Math. Phys. {\bf 79} (1981) 327;\\
            M. R. Douglas and S.H. Shenker, Nucl. Phys. {\bf B335} (1990)
635;\\
            D. Gross and A. Migdal, Phys. Rev. Lett. {\bf 64} (1990) 127;\\
            E. Brezin and V. Kazakov, Phys. Lett. B236 (1990) 144.
\bibitem{3} V.A. Kazakov, Phys. Lett. {\bf A119} (1986) 140;\\
             D.V. Boulatov and V.A. Kazakov, Phys. Lett. {\bf B186} (1987) 379.
\bibitem{3a} For a review see: C. Itzykson and J.-M Drouffe, ``Statistical
Field
Theory'', (Cambridge
University Press, 1989).
\bibitem{4a}C. Baillie and D. Johnston, Mod. Phys. Lett. {\bf A7}, 1519
            (1992); Phys. Lett. {\bf B286}, 44, (1992).
\bibitem{4} Z. Burda and J. Jurkiewicz, Acta Physica Polonica {\bf B20} (1989)
949.
            J. Jurkiewicz, A. Krzywicki, B. Petersson and B. Soderberg, Phys.
Lett. {\bf B213}             (1988) 511;\\
            R. Ben-Av, J. Kinar and S. Solomon, Nucl. Phys. {\bf B ( Proc.
Suppl.) 20}
            (1991) 711;\\
            S. Catterall, J. Kogut and R. Renken, Phys. Rev. {\bf D45},
            2957, (1992); Phys. Lett. {\bf B292}, 277, (1992);\\
            J. Ambj\o rn, B. Durhuus, T. Jonsson and G. Thorleifsson,
            Nucl. Phys. {\bf B398}, 569, (1993).
\bibitem{4b}M. Bowick, M. Falcioni, G. Harris and E. Marinari, ``Two
            Ising Models Coupled to 2 Dimensional Gravity'', SU-HEP-4241-556,
hep-th/9310136;
            ``Critical Slowing Down of Cluster Algorithms for Ising Models
Coupled
            to 2-d Gravity'', SU-HEP-4241-560, hep-lat/9311036.
\bibitem{6}D. Espriu, M. Gross, P.E.L. Rakow and J.F. Wheater, Nucl. Phys. {\bf
B265 (FS15)} (1986) 92;\\
             J.B. Zhang and D.R. Ji, Phys. Lett. {\bf A151} (1990) 431.
\bibitem{Dot} V. Dotsenko and V. Dotsenko, Adv. Phys. {\bf 32} (1983) 129.
\bibitem{7} W. Janke, M. Katoot and R. Villanova, Phys. Lett. {\bf B315} (1993)
412.
\bibitem{8} M.E. Agishtein and C.F. Baillie, Mod. Phys. Lett. {\bf A6} (1991)
1615.
\bibitem{9} M.E. Agishtein and A.A. Migdal, Nucl. Phys. {\bf B350} (1991) 690.
\bibitem{10} H. Kawai, Y. Kitazawa and M. Ninomiya, Nucl. Phys. {\bf B393}
(1993) 290.
\bibitem{11} H. Kawai, N. Kawamoto and Y. Watakibi, Phys. Lett. {\bf B306}
(1993) 19.
\bibitem{5a} S.F. Edwards and P.W. Anderson, J. Phys. {\bf F5} (1975) 965;\\
             M. Mezard, G. Parisi and M. Virasoro, ``Spin Glass Theory and
Beyond''
             World Scientific, 1987;\\
             K. Binder and A.P. Young, Rev. Mod. Phys. {\bf 58} (1986) 801.
\bibitem{12} D. Johnston, Phys. Lett. {\bf B277} (1992) 405.
\bibitem{13}V.I. Dotsenko and V.A. Fateev, Nucl. Phys. {\bf B240} (1984) 312,
{\bf B251} (1985) 691.
\bibitem{14} G. Harris, ``Percolation on Strings and the Coverup of the
C=1 Disaster'', Syracuse preprint SU-HEP-4241-555, hep-th/9310137.
\bibitem{Wh} J. Wheater, Phys. Lett. {\bf B198} (1987) 373;\\
             J. McCarthy, Nucl. Phys. {\bf B275} (1986) 421.
\end{thebibliography}
\end{document}